\documentclass[epj]{svjour}
\usepackage[sort&compress,numbers]{natbib}
\usepackage{amssymb}
\usepackage{graphicx}
\usepackage{geometry}
\usepackage[dvipsnames,table]{xcolor}
\usepackage{slashed}
\usepackage{hyperref} 
\usepackage{xspace}
\usepackage[tight]{subfigure}
\usepackage{amsmath}
\usepackage{amsfonts}
\usepackage{mathrsfs}
\usepackage{comment}
\usepackage{afterpage}
\usepackage{verbatim}
\usepackage{booktabs}
\usepackage{array}
\usepackage[title,titletoc]{appendix}
\usepackage{tabularx}
\newcommand{\stiny}[1]{{\scalebox{0.6}{#1}}}
\newcolumntype{C}[1]{>{\centering\arraybackslash}p{#1}}

\def\reffi#1{\mbox{Fig.~\ref{#1}}}
\def\reffis#1{\mbox{Figs.~\ref{#1}}}

\newcommand{\newc}{\newcommand}
\newc{\beq}{\begin{equation}}
\newc{\eeq}{\end{equation}}
\newc{\bit}{\begin{itemize}}
\newc{\eit}{\end{itemize}}
\newc{\ben}{\begin{enumerate}}
\newc{\een}{\end{enumerate}}
\newc{\bce}{\begin{center}}
\newc{\ece}{\end{center}}
\newc{\bfi}{\begin{figure}}
\newc{\efi}{\end{figure}}

\newcommand{\rT}{{\mathrm{T}}}

\newcommand{\rL}{{\mathrm{L}}}

\newcommand{\GeV}{\ensuremath{\,\text{GeV}}\xspace}

\newcommand{\Pj}{\ensuremath{\text{j}}\xspace}
\newcommand{\Pp}{\ensuremath{\text{p}}}
\newcommand{\Pe}{\ensuremath{\text{e}}\xspace}
\newcommand{\Pb}{\ensuremath{\text{b}}\xspace}

\newcommand{\Pt}{\ensuremath{\text{t}}\xspace}

\newcommand{\PW}{\ensuremath{\text{W}}\xspace}
\newcommand{\PZ}{\ensuremath{\text{Z}}\xspace}

\newcommand{\Mt}{\ensuremath{m_\Pt}\xspace}

\newcommand{\GZ}{\ensuremath{\Gamma_\PZ}\xspace}

\newcommand{\madgraph}{{\sc MadGraph}\xspace}

\newcommand{\phantommc}{{\sc Phantom}\xspace}

\newcolumntype{.}{D{.}{.}{-1}}
\newcolumntype{d}[1]{D{.}{.}{#1}}
\colorlet{tableoverheadcolor}{gray!37.5}
\colorlet{tableheadcolor}{gray!25}
\colorlet{tablerowcolor}{gray!12.5}

\def\draftdate{\relax}
\def\mda{\relax}
\def\mua{\relax}
\def\mla{\relax}
\def\draft{
\def\thtystars{******************************}
\def\sixtystars{\thtystars\thtystars}
\typeout{}
\typeout{\sixtystars**}
\typeout{* Draft mode!
         For final version remove \protect\draft\space in source file *}
\typeout{\sixtystars**}
\typeout{}
\def\draftdate{\today}
\def\mua{\marginpar[\boldmath\hfil$\uparrow$]%
                   {\boldmath$\uparrow$\hfil}\color{black}%
                    \typeout{marginpar: $\uparrow$}\ignorespaces}
\def\mda{\color{red}\marginpar[\boldmath\hfil$\downarrow$]%
                   {\boldmath$\downarrow$\hfil}%
                    \typeout{marginpar: $\downarrow$}\ignorespaces}
\def\mla{\marginpar[\boldmath\hfil$\rightarrow$]%
                   {\boldmath$\leftarrow $\hfil}%
                    \typeout{marginpar: $\leftrightarrow$}\ignorespaces}
\def\Mua{\marginpar[\boldmath\hfil$\Uparrow$]%
                   {\boldmath$\Uparrow$\hfil}\color{black}%
                    \typeout{marginpar: $\uparrow$}\ignorespaces}
\def\Mda{\color{red}\marginpar[\boldmath\hfil$\Downarrow$]%
                   {\boldmath$\Downarrow$\hfil}%
                    \typeout{marginpar: $\downarrow$}\ignorespaces}
\def\Mla{\marginpar[\boldmath\hfil\textcolor{red}{$\Rightarrow$}]%
                   {\boldmath\textcolor{red}{$\Leftarrow $}\hfil}%
                    \typeout{marginpar: $\leftrightarrow$}\ignorespaces}
\overfullrule 5pt
\oddsidemargin 15mm
\marginparwidth 29mm
}

\newcommand{\mc}{\mathcal}
\let\nnb\notag

\let\Mw\MW
\let\Gw\GW
\let\Mz\MZ
\let\Gz\GZ

\let\as\alphas
\newcommand{\pt}[1]{p_{\rT,{#1}}}

\newcommand{\rf}[1]{ref.~\cite{#1}}
\newcommand{\Rf}[1]{Ref.~\cite{#1}}
\newcommand{\rfs}[1]{refs.~\cite{#1}}

\begin{document}
\title{
  Polarized $\PZ$ bosons from the decay of a Higgs boson  produced in association
  with two jets at the LHC
}


     \author{Ezio Maina\footnote{\label{a}} \and Giovanni Pelliccioli\footnote{}}
     \institute{\textsuperscript{a}University of Torino, Department of Physics and INFN, Sezione di Torino, via Pietro Giuria 1, 10125 Torino (Italy) \\
      \textsuperscript{b}University of W\"urzburg, Institute for Theoretical Physics and Astrophysics, Emil-Hilb-Weg 22, 97074 W\"urzburg (Germany)}

\abstract{
  Investigating the polarization of weak bosons provides an important probe of the scalar and gauge
  sector of the Standard Model. This can be done in the Higgs decay to four leptons, whose
  Standard-Model leading-order amplitude enables to generate polarized observables from unpolarized
  ones via a fully-differential reweighting method. We study the $\PZ$-boson polarization from the decay of
  a Higgs boson produced in association with two jets, both in the gluon-fusion and in the
  vector-boson fusion channel. We also address the possibility of extending the results of this work
  to higher orders in perturbation theory.
}
\authorrunning{\emph{E. Maina, G. Pelliccioli}}
\titlerunning{\emph{
}}
\maketitle

\section{Introduction}\label{intro}
The discovery of the Higgs boson \cite{Chatrchyan:2012xdj,Aad:2012tfa} has enabled a large
number of tests of the Standard Model (SM) scalar and gauge sectors.

In the SM, electroweak bosons ($\PW$ and $\PZ$) acquire their mass and, as a consequence,
an additional longitudinal polarization, through their coupling to the Higgs field.
Therefore the interplay between the Higgs boson
and weak-boson polarizations represents a crucial probe for the SM electroweak symmetry
breaking mechanism as well as an ideal framework for searches of possible modifications
due to new-physics effects.

The gold-plated channel for these searches is given by vector-boson scattering (VBS),
as in the high-energy regime the unitarity-violating behaviour of the scattering among
longitudinal bosons is regularized by the inclusion of Higgs exchange contributions
\cite{Dicus:1992vj,LlewellynSmith:1973yud,Veltman:1976rt}.
Studying the Higgs coupling to polarized weak bosons can provide relevant
insight also at the {Higgs-boson} resonance itself \cite{Brehmer:2014pka},
in particular with the aim of discriminating between the SM and models
with modified Higgs sectors. Since {vector-boson polarizations} are best accessed through
the angular distribution of their decay products, the cleanest channel is the Higgs
decay to four charged leptons.

The {Higgs-boson} decay into four charged leptons has been widely studied by the
ATLAS and CMS experimental collaborations with Run-1 (7 and 8 TeV) and Run-2 data (13 TeV),
with the purpose of determining the {Higgs-boson} mass
\cite{Chatrchyan:2013mxa,Aad:2014eva,Sirunyan:2017exp}
and width
\cite{Khachatryan:2014iha,Khachatryan:2015mma,Sirunyan:2019twz},
its spin and parity properties
\cite{Chatrchyan:2013mxa,Khachatryan:2014kca,Aaboud:2017vzb,Aad:2020mkp},
the inclusive and differential cross-sections
\cite{Aad:2014tca,Khachatryan:2015yvw,Aaboud:2017oem,Aaboud:2017vzb,Aad:2020mkp,ATLAS:2020wny,Sirunyan:2021rug},
and the Higgs coupling to weak bosons in the on-shell and off-shell regions
\cite{Khachatryan:2014kca,Aaboud:2017vzb,Sirunyan:2017tqd,Aaboud:2018puo,Sirunyan:2019twz,Sirunyan:2021fpv}.
So far, all measurements are compatible with the SM predictions.
This decay channel features a large signal-to-background ratio, thanks to
the possibility to fully reconstruct the decay products and to small background
contributions.

The four-lepton {Higgs-boson} decay has been extensively investigated
also from a theoretical point of view,
providing a very precise characterisation of the scalar boson properties.
The SM decay of a Higgs boson into four leptons is known perturbatively up to 
\sloppy next-to-leading order (NLO)
in the electroweak (EW) coupling \cite{Bredenstein:2006rh}
and has been matched to QED parton-showers \cite{Boselli:2015aha}.
This decay channel has been computed at EW NLO also for beyond-the-SM theories with
modified Higgs sectors  
\cite{Altenkamp:2018bcs,Altenkamp:2018hrq,Kanemura:2019kjg},
and has been investigated within an effective-field-theory framework 
\cite{Artoisenet:2013puc,Boselli:2017pef,Brivio:2019myy}.
Spin effects, angular and energy correlations have been widely studied
\cite{Soni:1993jc,Chang:1993jy,Skjold:1993jd,Arens:1994wd,Buszello:2002uu,Choi:2002jk,Hagiwara:2009wt,Gao:2010qx,Bolognesi:2012mm,Berge:2015jra}.

At the Large Hadron Collider (LHC), the Higgs boson is mostly produced
in the gluon-fusion (GGF) and in the vector-boson-fusion (VBF) channels.
Experimental analyses can be either inclusive or exclusive in the production
mode of the Higgs boson, depending on the specific target of the investigation.
For the purposes of this work, we only consider the production of a Higgs boson
in association with two jets. In this channel the GGF and VBF contributions are comparable.
The SM prediction for an on-shell Higgs boson produced in GGF in association with
two jets is known at leading order (LO) in QCD, including also dependence on the top-quark
mass \cite{DelDuca:2001eu,DelDuca:2001fn,Greiner:2016awe}. Predictions in the
large-$\Mt$ limit are known up to NLO QCD \cite{Cullen:2013saa}.
The full mass dependence has also been studied in the presence
of high-energy jets \cite{Andersen:2018kjg}. 
The predictions for VBF are known at NLO QCD+EW accuracy \cite{Ciccolini:2007jr},
and up to $\rm N^3LO$ QCD \cite{Dreyer:2016oyx} in the
structure-function approximation \cite{Han:1992hr}.
It has been shown \cite{DelDuca:2001eu,DelDuca:2001fn}
that for an on-shell Higgs produced in GGF in association with two jets,
the large-$\Mt$ approximation gives a very good description of
the loop-induced SM process with full $m_t$ dependence,
of order $\mc O (\as^4\alpha)$, provided that the transverse-momenta of jets
and the {Higgs-boson} mass are smaller than the top-quark mass.
The logarithmic structure of Higgs production in association with up to two jets
and the kinematic configurations where finite top-mass effects become relevant
are also known \cite{Buschmann:2014twa,Buschmann:2014sia}.
The large-$\Mt$ approximation works well also in the presence of a
two-jet system with large invariant mass \cite{DelDuca:2001eu,DelDuca:2001fn,Campbell:2012am},
which is the typical VBF phase-space region.
The interference between VBF and GGF signals has been proved to be negligible
\cite{Andersen:2007mp}.

The phenomenology of weak-boson polarizations has been widely
investigated both in experimental analyses and in theoretical
studies.

{\color{black}The ATLAS and CMS collaborations have measured final-state vector-boson polarizations
(typically using the leptonic decay channel) in several multi-boson processes in 8 and 13 TeV hadronic collisions, including
$V$+jets \cite{Chatrchyan:2011ig,ATLAS:2012au,Khachatryan:2015paa,Aad:2016izn},
di-boson production \cite{Aaboud:2019gxl} and
vector-boson scattering \cite{Sirunyan:2020gvn}. Polarizations have also been
measured for $\PW$ bosons produced in top-quark decays \cite{Aaboud:2016hsq,Khachatryan:2016fky,Aad:2020jvx},
as well as for $V$ bosons produced in Higgs-boson decays, ${\rm h}\rightarrow VV^*$ \cite{Aaboud:2018jqu,Bruni:2019xwu}.}
Enhanced sensitivity to polarizations is expected in the forthcoming
high-luminosity and high-energy LHC runs \cite{CMS-PAS-FTR-18-014,Azzi:2019yne}.

From the theory side, the $\PW$-boson polarization at the LHC has been studied in
\rf{Bern:2011ie} in the absence of lepton cuts. 
Realistic selection cuts has been introduced in \rf{Stirling:2012zt} both
in $V+{}$jets and in other multi-boson production processes.
The interference between amplitudes for different polarizations has
been investigated in \rf{Belyaev:2013nla}.

In \rf{Ballestrero:2017bxn}
a simple and natural method to define cross sections
corresponding to vector bosons of definite polarization has been proposed.
This method has been applied to study the polarization of $\PW$
and $\PZ$~bosons in vector-boson scattering at LO
\cite{Ballestrero:2017bxn,Ballestrero:2019qoy,Ballestrero:2020qgv}
and in di-boson production at NLO QCD
\cite{Denner:2020bcz,Denner:2020eck}
with purely leptonic final states.
Polarizations in $\PW^\pm \PZ$ production, including NLO QCD and EW
corrections have been {analyzed} in  \rfs{Baglio:2018rcu,Baglio:2019nmc}.
Recently \madgraph has introduced the possibility of generating
polarized amplitudes \cite{BuarqueFranzosi:2019boy}.
\Rf{Cao:2020npb} has suggested that a study of {vector-boson} polarizations in
gluon-induced $\PZ\PZ$ production could be sensitive to the
$\PZ\Pt\bar{\Pt}$ coupling.

A study of the vector-boson polarization effects at the Higgs resonance in VBF has
been performed in the $\PW^+\PW^-$ channel, using effective Higgs couplings to longitudinal
and transverse bosons and including a general dimension-six EFT interpretation
\cite{Brehmer:2014pka}.
The possibility of measuring the coupling of the Higgs to polarized bosons is studied in \rf{Bruni:2019xwu}.

In this work we perform a phenomenological study of polarized electroweak bosons from
the decay of a SM Higgs boson produced in association with two jets
at the LHC. In \rf{Maina:2020rgd} a first assessment of polarized bosons from
Higgs decays has been performed in the LO production of a Higgs in gluon
fusion, that is, for a Higgs with vanishing transverse momentum.
In this paper we consider the case in which the Higgs boson is produced in association with
two jets, both in VBF and in GGF. Although the focus
is on the Higgs signals, a number of comments are made concerning the
impact of the QCD background and of the pure electroweak contributions to VBF.

The phenomenological analysis is limited to the four-lepton Higgs decay, with
two pairs of opposite-charge leptons and different flavours. Despite a very small
cross-section, due to the small branching ratio, the considered decay channel has
two advantages. First, it enables the complete reconstruction of the {Higgs-boson}
kinematics, thanks to the absence of neutrinos in the final state. Second, it allows
for an unambiguous determination of the kinematics of each $\PZ$-boson, thanks to the
different lepton flavours.

VBF and GGF Higgs production with two jets has already been widely
investigated \cite{Klamke:2007cu,Andersen:2008gc,Andersen:2012kn,Demartin:2014fia},
mostly with the purpose of finding kinematic regimes and observables that
discriminate between the two signals. In this work, however, we do not aim to 
separate the Higgs signals, instead, we address the possibility to perform
a polarization study of vector bosons in Higgs decay, rather
independently of the {Higgs-production} mechanism. Given the very small fraction
of Higgs bosons decaying into four charged leptons, being able to sum
over different production channels is crucial to enhance the experimental
sensitivity to the polarization structure.

As a last comment, we choose to work in the large-$\Mt$ limit for the
GGF signal. Given the purposes of this work, it represents
a satisfactory approximation.

This paper is organised as follows. In Sect.~\ref{reweighting} we show the details
of the matrix-element reweighting method that we use to generate polarized cross-sections
and distributions.
The fiducial setup we employ for numerical simulations is given in Sect.~\ref{sec:setup}.
In Sect.~\ref{sec:valid} we validate the reweighting technique comparing its results
with those obtained directly simulating polarized events \cite{Ballestrero:2017bxn}.
The polarized results for VBF and GGF signals are shown and discussed in
Sect.~\ref{sec:results}. In Sect.~\ref{sec:concl} we draw our conclusions.

\section{Matrix-element reweighting}\label{reweighting}
\begin{figure}[h!]
  \centering
  \hspace*{1cm}\includegraphics[scale=0.45]{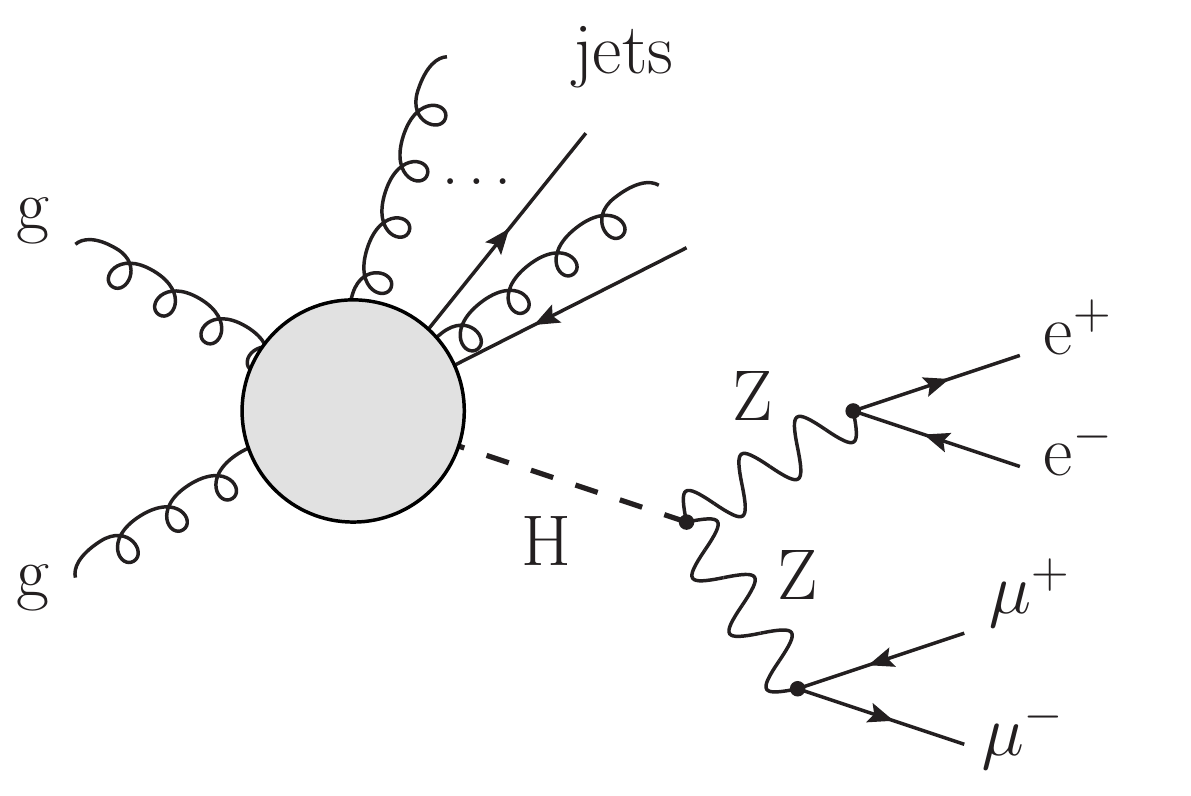}
  \caption{General structure of the Standard Model amplitude for the production of
  a Higgs boson (decaying into four charged leptons) in association with jets.}\label{fig:feynH}
\end{figure}

%
%
In Fig.~\ref{fig:feynH} we show the general structure of Higgs-mediated amplitudes
contributing to a reaction producing the Higgs-decay products (four charged-leptons)
and an arbitrary number of jets.
{\color{black}
 
  We notice that these amplitudes (which we call \emph{signal contributions}) are only a sub-set of all
  amplitudes for the production of four charged leptons in association with jets.
  Therefore, they are not gauge invariant \emph{per s\`e}. Nonetheless, we have checked numerically that
  with a sharp but realistic cut on the invariant mass of the four leptons, \emph{e.g.} $|M_{4\ell}-M_{\rm h}|<5\GeV$,
the $s$-channel Higgs contributions are largely dominant and sufficient to describe the full-matrix-element calculation up
  to few-permille effects.
}

We observe that for this work it is important that the Higgs-decay products
can be unambiguously identified and distinguished from the additional jets produced
in association with the Higgs boson in order to minimize the impact of irreducible backgrounds
and reconstruction effects.

Provided we define { vector-boson polarizations} in the Higgs rest frame,
we can easily parametrize the Higgs-decay amplitude as a sum of polarized and interference
terms \cite{Maina:2020rgd}:\\[-0.2cm]
\begin{eqnarray}\label{eq:master}
  |\mc A^{\rm SM}_{h\rightarrow \PZ\PZ}|^2
  &=&
  \Big[|\mc A_{\rL\rL}|^2 +
  |\mc A_{++}|^2 
  +  |\mc A_{--}|^2 \nnb\\
  && +2\, {\rm Re}(\mc {A_{\rL\rL}}^{\!*}\mc A_{++})
     +2\, {\rm Re}(\mc {A_{\rL\rL}}^{\!*}\mc A_{--})\nnb\\
     &&   +2\, {\rm Re}(\mc {A_{++}}^{\!*}\mc A_{--})\,\Big]
\end{eqnarray}
where the first three terms in the sum correspond to
{\color{black}a definite polarization state $\lambda$ for both bosons,
longitudinal ($\lambda={\rm L}$), right-handed ($\lambda=+$) and left-handed ($\lambda=-$),}
while the fourth and fifth terms are contributions
from longitudinal-transverse interference, and the
last one comes from left-right interference. Note that,
since polarization vectors are defined in the {Higgs-boson}
rest frame there are no mixed-state contributions,
\emph{i.e.} $\mc A_{\lambda\lambda'}=0$ for $\lambda\neq\lambda'$.
The analytic expression for each term of the sum
of Eq.~\ref{eq:master} reads
\begin{eqnarray}\label{ll}
  |\mc A_{\rL\rL}|^2 &=&
  4\,K^2\,P(Q_1)\,P(Q_2)\,\\
  &\times&  \left(c_L^2+c_R^2\right)^2\,\sin^2\theta^*_{1}\, \sin^2\theta^*_{2}\nnb\\[-0.7cm]\nnb
\end{eqnarray}
\begin{eqnarray}
  |\mc A_{++}|^2 &=& \label{pp}
  \,P(Q_1)\,P(Q_2)\,\\
  &\times& \Big[\,c_L^4\,(1+\cos\theta^*_{1})^2\,(1+\cos\theta^*_{2})^2\nnb\\
    && + c_R^4\,(1-\cos\theta^*_{1})^2\,(1-\cos\theta^*_{2})^2 \nnb\\[0.12cm]
    && + c_R^2c_L^2\,(1+\cos\theta^*_{1})^2\,(1-\cos\theta^*_{2})^2 \nnb\\
    && + c_R^2c_L^2\,(1-\cos\theta^*_{1})^2\,(1+\cos\theta^*_{2})^2\, \Big]\nnb\\[-0.7cm]\nnb
\end{eqnarray}
\begin{eqnarray}
  |\mc A_{--}|^2 &=& \label{mm}
  \,P(Q_1)\,P(Q_2)\,\\
  &\times& \Big[\,c_L^4\,(1-\cos\theta^*_{1})^2\,(1-\cos\theta^*_{2})^2\nnb\\
    && + c_R^4\,(1+\cos\theta^*_{1})^2\,(1+\cos\theta^*_{2})^2 \nnb\\[0.12cm]
    && + c_R^2c_L^2\,(1+\cos\theta^*_{1})^2\,(1-\cos\theta^*_{2})^2 \nnb\\
    && + c_R^2c_L^2\,(1-\cos\theta^*_{1})^2\,(1+\cos\theta^*_{2})^2\, \Big]\nnb\\[-0.7cm]\nnb
\end{eqnarray}
\begin{eqnarray}
2\, {\rm Re}(\mc {A_{\rL\rL}}^{\!*}\mc A_{++})
  &=& \label{lm}
  \,-4\,K\,P(Q_1)\,P(Q_2)\,\\
  &\times& \Big[\,c_L^4\,(1+\cos\theta^*_{1})\,(1+\cos\theta^*_{2})\nnb\\    
    && + c_R^4\,(1-\cos\theta^*_{1})\,(1-\cos\theta^*_{2}) \nnb\\[0.12cm]
    && - c_R^2c_L^2\,(1+\cos\theta^*_{1})\,(1-\cos\theta^*_{2}) \nnb\\
    && - c_R^2c_L^2\,(1-\cos\theta^*_{1})\,(1+\cos\theta^*_{2})\, \Big]\nnb\\
   &\times& \sin\theta^*_{1}\, \sin\theta^*_{2}\,\cos\Delta\phi^*_{12}\nnb\\[-0.7cm]\nnb
\end{eqnarray}
\begin{eqnarray}
2\, {\rm Re}(\mc {A_{\rL\rL}}^{\!*}\mc A_{--})
  &=& \label{lp}
  \,-4\,K\,P(Q_1)\,P(Q_2)\,\\
  &\times& \Big[\,c_L^4\,(1-\cos\theta^*_{1})\,(1-\cos\theta^*_{2})\nnb\\    
    && + c_R^4\,(1+\cos\theta^*_{1})\,(1+\cos\theta^*_{2}) \nnb\\[0.12cm]
    && - c_R^2c_L^2\,(1+\cos\theta^*_{1})\,(1-\cos\theta^*_{2}) \nnb\\
    && - c_R^2c_L^2\,(1-\cos\theta^*_{1})\,(1+\cos\theta^*_{2})\, \Big]\nnb\\
  &\times& \sin\theta^*_{1}\, \sin\theta^*_{2}\,\cos(\Delta\phi^*_{12})\nnb\\[-0.7cm]\nnb
\end{eqnarray}
\begin{eqnarray}
 2\, {\rm Re}(\mc {A_{++}}^{\!*}\mc A_{--})
  &=& \label{pm}
  \,2\,P(Q_1)\,P(Q_2)\,\\
  && \hspace*{-1cm} \times\,\left(c_L^2+c_R^2\right)^2\,\sin^2\theta^*_{1}\,
  \sin^2\theta^*_{2} \,\cos(2\Delta\phi^*_{12})\,.\nnb
\end{eqnarray}
In Eqs.~\ref{ll}--\ref{pm} we have defined the $P(Q_i)$ propagator factor as\\[-0.15cm]
\beq
P(Q_i)\,=\, \frac{2\,g_{\stiny{\rm hZZ}} \,Q_{i}^2}{\left(Q_{i}^2-\Mz^2\right)^2+\GZ^2\Mz^2}\,,
\eeq
depending on the Higgs-to-gauge-boson coupling ($g_{\rm \stiny{\rm hZZ}}$),
the $\PZ$ pole mass and width, and the invariant mass of the $i$-th $\PZ$ boson ($Q_i$).
We have also introduced the factor\\[-0.2cm]
\beq
K = \frac{Q^2-Q_1^2-Q_2^2}{2\,Q_1\,Q_2}\,,
\eeq
where $Q$ is the {Higgs-boson} invariant mass.
The $c_L,c_R$ parameters represent the SM left- and right-chirality
coupling of the $\PZ$ boson to massless leptons. The variables $\theta^*_{1(2)},\phi^*_{1(2)}$
are the positively-charged-lepton decay angles computed in the corresponding $\PZ$-boson
rest frame, with respect to the boson flight direction in the Higgs rest frame, and
$\Delta\phi^*_{12}=\phi^*_1-\phi^*_2$.

Instead of considering the left-handed and right-handed contributions
separately, we combine them in a single transverse (T) contribution
which includes also the left-right interference term,\\[-0.2cm]
\beq\label{ATT}
|\mc A_{\rT\rT}|^2 
  =
  |\mc A_{++}|^2 
  +  |\mc A_{--}|^2 + 2\, {\rm Re}(\mc {A_{++}}^{\!*}\mc A_{--})\,.
  \eeq
The definition of the transverse mode as a coherent sum of the left
and right modes minimizes the interference effects, which
now come only from longitudinal-transverse terms in Eqs.~\ref{lm}-\ref{lp}.
Moreover, it is known \cite{Maina:2020rgd} that in the Higgs decay to four charged leptons
the longitudinal-transverse interference is only a few percent of the total result, much smaller
than the interference between the right and left amplitudes included in Eq.~\ref{ATT}.
Introducing the following longitudinal or transverse weights,\\[-0.2cm]
\beq\label{eq:weight}
w_{\lambda} = \frac{
  \big|\mc A_{\lambda\lambda}\big|^2}{\big|\mc A^{\rm SM}_{h\rightarrow \PZ\PZ}\big|^2}
\,,\quad \lambda = \rL,\rT\,,
\eeq
it is possible to compute any polarized distribution, with arbitrary cuts on the lepton
kinematics, multiplying the weight of each unpolarized
event by the factor in Eq.~\ref{eq:weight}. In this way it becomes unnecessary to generate
separately the individual polarized contributions. All the subtleties related to
sampling events with negative weights are avoided.  

We stress that this procedure is accurate only because of the special
(and relatively simple) Higgs-decay analytic structure. Applying the same
approach to general multi-boson processes would require building weights
that depend on non-factorized amplitudes, which would be much more time-consuming
and equivalent to directly generate events with polarized amplitudes.

This method is designed for the LO Higgs decay into four leptons. However, it could
be extended to NLO, provided that it is possible to disentangle polarized contributions
to the $\PZ$ boson propagators. The NLO QCD corrections would only affect the production
mechanism, therefore this entire formalism can be extended with no modifications to this
perturbative order.
{However, we do not include QCD radiative corrections to the specific production
processes that are considered in this paper, as this would not affect the
main results of this work.
The extension of this reweighting method} to NLO EW corrections is more involved, in particular
for the virtual contributions. These corrections are of the order of 2\% \cite{Bredenstein:2006rh}.
{The description of weak boson  polarizations is not trivial if EW corrections
are included, since a standard narrow-width or double-pole approximation \cite{Denner:2000bj} is not viable
for vector bosons from a 125-GeV {Higgs-boson} decay.
Below the $2M_{\rm V}$ threshold the NLO EW corrections can be computed via
an improved-Born approximation \cite{Denner:2019vbn}, which is only valid
for $M_{\rm VV}\approx 2 M_{\rm V}-n\Gamma_{\rm V}$, with $n=2\div3$.
However, a rigorous description of {weak boson} polarizations in Higgs decay at NLO EW
is possible in the single-pole approximation, namely projecting on mass shell only one
of the two vector bosons. This is discussed in Sect.~\ref{sec:SPA}.}

\section{Single-pole approximation}\label{sec:SPA}
In this section we briefly address the possibility of studying $\PZ$ boson polarizations
in Higgs decay including NLO EW effects.
The presence of non-factorizable corrections both in the real and the virtual contributions
makes the separation of polarizations of off-shell $\PZ$ bosons not well defined, as
several diagram topologies do not feature two $s$-channel $\PZ$-boson propagators.
Such diagrams cannot be simply dropped as the result would not be gauge invariant.
The pole approximation \cite{Stuart:1991xk,Aeppli:1993rs} can help in addressing this issue.
Since on the Higgs resonance one of the two $\PZ$ bosons is off its mass shell,
it is only possible to project on-shell one of the two bosons. 
This method consists in:
\begin{itemize}
\item selecting only diagrams which feature at least a resonant $\PZ$ boson decaying into two leptons,
\item projecting the momenta of the $s$-channel $\PZ$-boson decay products such that their sum gives
  the momentum of an on-shell $\PZ$ momentum,
\item computing the numerator of the amplitude with the on-shell-projected kinematics, while
  evaluating the denominator with the original kinematics,
  to retain the Breit-Wigner modulation of an off-shell $\PZ$ boson.
\end{itemize}
The method is accurate in the vicinity of the resonance pole mass
\cite{Stuart:1991xk,Aeppli:1993rs,Denner:2019vbn},
and is characterized by an intrinsic uncertainty of order $\mc O (\Gz/\Mz) \approx \mc O (\alpha)$.
This procedure requires a careful treatment of non-factorizable QED corrections
\cite{Denner:1997ia,Dittmaier:2015bfe,Denner:2019vbn},
because of spurious infrared singularities that arise when setting the $\PZ$-boson
momentum on its mass shell.
Furthermore, the presence of photon contributions in the virtual corrections makes
it essential to constrain the decay leptons to have an invariant
mass that is as close as possible to the $\PZ$ pole mass. This requirement has 
the advantage of making the $\PZ$-resonant contributions more dominant over the
non-resonant ones, giving a more accurate pole-approximated description of the process.
This induces the loss of signal events, due to the tighter constraint on the
two leptons decaying from the almost-on-shell $\PZ$ boson.
We have calculated that with realistic kinematic selections (see LEP setup of Sect.~\ref{sec:setup}),
the additional constraint $|M_{\Pe^+\Pe^-}-\Mz|<5\GeV$ reduces the VBF total cross-section
by 70\%.
Most of the effect comes from the fact that only half of the events contain an
almost-on-shell $\Pe^+\Pe^-$ pair. However, one should include also the contribution
from almost-on-shell $\mu^+\mu^-$ pairs.

The choice of the specific on-shell projection is not unique
\cite{Denner:2000bj,Ballestrero:2017bxn,Ballestrero:2019qoy,Denner:2019vbn},
and introduces an artificial modification of the momenta, which depends 
{on the physical quantities that} one chooses to conserve.
In Higgs decay, the choice of such a projection is even more delicate than in
other processes, as changing the momenta of one $\PZ$ boson could induce a shift
in the total momentum of the Higgs boson itself, resulting in a bad phase-space sampling
in the region of the Higgs resonance and in large discrepancies w.r.t. the
full calculation. Therefore, an essential requirement for the on-shell projection
is to preserve the total momentum of the four leptons.
A viable projection for the process $\rm h \rightarrow \PZ_1(\rightarrow \Pe^+\Pe^-)\,\PZ^*_2(\rightarrow \mu^+\mu^-)$ preserves:
\begin{itemize}
\item the spatial direction of $\Pe^\pm$ in the rest frame of the $\Pe^+\Pe^-$ system,
\item the spatial direction of $\mu^\pm$ in the rest frame of the $\mu^+\mu^-$ system,
\item the spatial direction of the $\Pe^+\Pe^-$ system in the rest frame of the four-lepton system,
\item the invariant mass of the $\mu^+\mu^-$ system,
\item the four-momentum of the four-lepton system.
\end{itemize}
This choice preserves the decay angles of the leptons in the corresponding
$\PZ$-boson rest frame, minimizing the effect of the pole approximation on
polarization-sensitive variables. This specific on-shell projection only works if
the invariant mass of the Higgs boson is larger than the sum $\Mz+M_{\mu^+\mu^-}$,
giving a decrease in the total cross-section, which is of the same order of magnitude as
the intrinsic uncertainty of the pole approximation.

This approach, despite some technical details which must be properly taken care of,
is expected to give a reliable description of the {Higgs-boson} decay into four leptons at NLO EW,
in the case where one of the two lepton pairs is close to the $\PZ$-boson
mass shell. The single-pole approximation allows to select in a gauge invariant way
only resonant contributions and therefore to reliably separate the polarizations of a single boson
at the amplitude level.
Since in the SM the Higgs couples to two weak bosons with like-wise polarization mode,
selecting the polarization mode of the on-shell $\PZ$ boson intrinsically gives important
information about the polarization of the off-shell boson, which can be then studied
by means of the usual angular observables of its decay leptons.

We conclude that the single-pole approximation represents a viable procedure
to extend the LO polarization studies on the Higgs resonance to higher orders,
in particular for NLO EW corrections. We leave this for future investigations.

\section{Setup}\label{sec:setup}
We now proceed to the results for the parton-level process
\beq
\Pp\, \Pp \rightarrow \Pj\, \Pj \,{\rm h}(\rightarrow\Pe^+\Pe^-\mu^+\mu^-)\,,
\eeq
at the LHC with 13 TeV centre-of-mass energy.
{\color{black}We have computed VBF and GGF signals at LO, the former simulated in the SM,
the latter in the large-$m_{\Pt}$ approximation, with the same numerical setup. A unit
CKM matrix is assumed in both processes.}
We use \texttt{NNPDF3.0} parton distribution functions (PDF)
\cite{Ball:2014uwa} computed at LO with $\as(\Mz)=0.118$, via the
{\scshape{LHAPDF}} interface \cite{Buckley:2014ana}.
The complex-mass scheme \cite{Denner:2005fg,Denner:2006ic} is understood
for the treatment of electroweak-boson masses and couplings in the SM.
The pole masses and widths of weak bosons and of the Higgs are set to the following values:
\begin{eqnarray}
  \Mw & = & 80.358  \GeV\,,\qquad\Gw=2.084\GeV\,,\nnb\\
  \Mz & = & 91.153\GeV\,,\qquad\,\,\Gz=2.494\GeV\,,\nnb\\
  M_{\rm h} & = & 125\GeV\,,\qquad\,\,\, \quad\Gamma_{\rm h}\,=\, 4.07{\,\rm MeV}\,.
\end{eqnarray}
The electroweak coupling $\alpha$ is computed in the $G_{\mu}$ scheme, with
the Fermi constant set to $G_\mu=1.16637 \cdot 10^{-5}\GeV^{-2}$.
For both GGF and VBF, we work in the five-flavour scheme, \emph{i.e.} including 
partonic processes with external massless $\Pb$-quarks. These contributions account for
less than 2\% of the total cross-section.

We use a dynamical factorization and renormalization scale,
\beq\label{eq:muF}
\mu_{\rm F} = \mu_{\rm R} = \sqrt{\pt{j_1}\pt{j_2}}\,,
\eeq
{\color{black}that is a typical choice for GGF production \cite{DelDuca:2001eu,DelDuca:2001fn}.}
We have verified that the typical scale choice used for
on-shell Higgs production \cite{Cacciari:2015jma} gives results for VBF which differ by a few percent (both at the
integrated and differential level) from those obtained with the dynamical choice in
Eq.~\ref{eq:muF}. 
{We stress that the scale choice does not affect the main results
of our work, as the reweighting procedure detailed in Sect.~\ref{reweighting} is independent of the Higgs-production
mechanism and its corresponding central-scale choice.}
We have employed two different setups.\\
The first setup (label INC) is inclusive in lepton kinematics,
and is used for validation purposes:
\begin{itemize}
\item[-] jets with minimum tranverse momentum $\pt{j}>25\GeV$ and maximum { pseudo-rapidity $|\eta_{j}|<4.5$};
\item[-] a two-jet system with rather large invariant mass, $M_{\Pj\Pj}>300\GeV$, large pseudo-rapidity separation, $|\Delta \eta_{\rm jj}|>2.5$, and
  such that $\eta_{j_1}\cdot\eta_{j_2}<0$;
\item[-] a four-lepton invariant mass close to the Higgs pole mass, $|M_{4\ell}-M_{\rm h}|<5\GeV$;
\item[-] two pairs of opposite-charge leptons of different flavours, with  $M_{\ell^+\ell^-}>10\GeV$, for $\ell=\Pe,\mu$.
\end{itemize}

The requirements on the leading-jet pair are somewhat milder than those
used for on-shell Higgs production in VBF \cite{Cacciari:2015jma} and slightly
stronger than those used in $\PZ\PZ$ scattering studies \cite{Denner:2020zit}.

The second setup (label LEP) includes, in addition to the ones of the INC setup, the
following cuts on the lepton kinematics:
\begin{itemize}
\item[-] for all charged leptons, $\pt{\ell^\pm}>20\GeV$, {$|\eta_{\ell^\pm}|<2.5$}.
\end{itemize}

Note that the effect of the lepton $p_{\rm T}$ and $\eta$ cuts is dramatic, as
they decrease the signal cross-section roughly by a factor of 10, with respect
to the INC setup.
These cuts are the same used in the study of polarized $\PZ$-boson scattering
performed in \rf{Ballestrero:2019qoy}.
In recent experimental results on Higgs decay to four charged leptons
\cite{Aad:2021ebo,Sirunyan:2021rug} the fiducial lepton cuts are slightly looser than
those used in this work, therefore we expect a milder effect on polarized results
in a realistic experimental analysis than in the present phenomenological setup.

For the simulation of the unpolarized VBF signal we have used the \phantommc Monte Carlo
\cite{Ballestrero:2007xq},
     {\color{black}which enables the selection of diagrams with an $s$-channel Higgs exchange
       from the complete set of tree-level, electroweak diagrams in the SM \cite{Ballestrero:2015jca}.}
The same process has been simulated also with \madgraph \cite{Alwall:2014hca},
with an agreement to better than 
0.5\% both in the total cross-section and in all analyzed differential
distributions.
The GGF signal has been simulated with \madgraph (version 2.7.3 \cite{BuarqueFranzosi:2019boy}),
in the large-$m_{\Pt}$ approximation \cite{Artoisenet:2013puc}, using the spin-correlated decay chain
for the $s$-channel Higgs boson.
We observe that the off-shell-ness of the Higgs boson is preserved both in \madgraph
and in \phantommc. The numerical integration is carried out by means of a Breit-Wigner
phase-space mapping restricted to the region $|M_{4\ell}-M_{\rm h}|<5\GeV$, which is also
the kinematic region determined by the selection cuts.

\begin{table}[tb]
\begin{center}
\renewcommand{\arraystretch}{1.3}
\begin{tabular}{c|ccc|cc}%
  &\multicolumn{3}{c|}{ setup INC } & \multicolumn{2}{c}{ setup LEP }  \\[0.5ex]
  \hline
  pol.  &  MC & Rew. & Leg. &  MC & Rew. \\[0.5ex]
  \hline
unp. & 91.68(7) & - & - & 8.566(5) & - \\[0.5ex]
LL & 53.79(4) & 53.76(5) & 53.8(2) & 5.507(4)  & 5.505(5) \\[0.5ex]   
TT & 37.90(3) & 37.89(4) & 37.8(1) & 3.055(9) &  3.057(3)\\[0.5ex]
\end{tabular}
\end{center}
\caption{Total cross-sections (in $10^{-6}\,$pb) in the INC and LEP
  setups, obtained with polarized amplitudes (MC)
  and with the matrix-element reweighting method (Rew.). In the INC setup
  also the results of Legendre-polynomial projections (Leg.) are shown.
  Numerical errors are given in parentheses.
}\label{tab11}
\end{table}

  \begin{figure*}[htb]
    \subfigure[$\cos\theta^*_{\Pe^+}$\label{thetastar_rew}]{\includegraphics[scale=0.44]{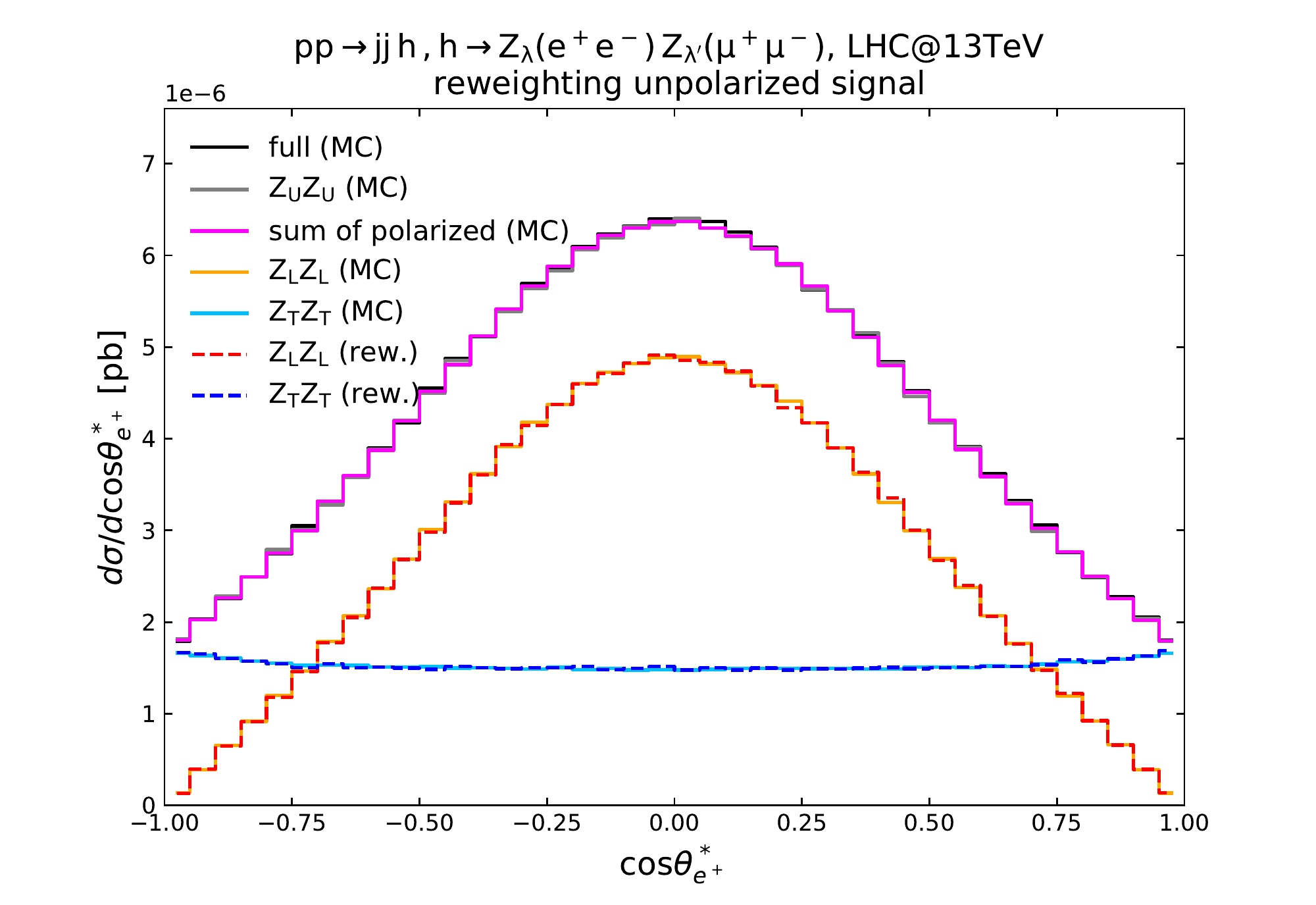}}
    \subfigure[$\Delta\phi_{\Pe^+\mu^+}$ \label{deltaphi_rew}]      {\includegraphics[scale=0.44]{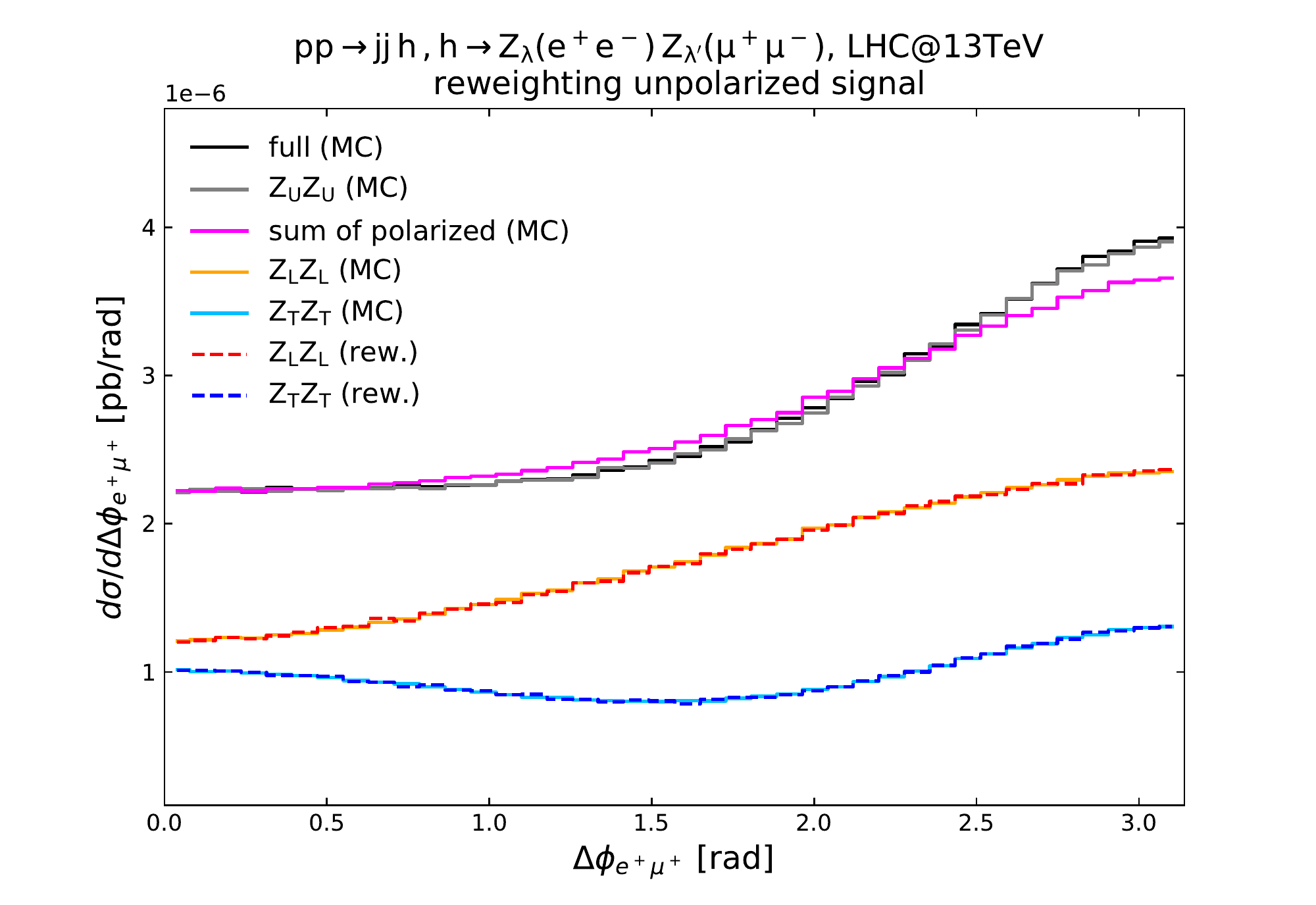}}
    \subfigure[$M_{\Pe^+\Pe^-}$\label{mee_rew}]      {\includegraphics[scale=0.44]{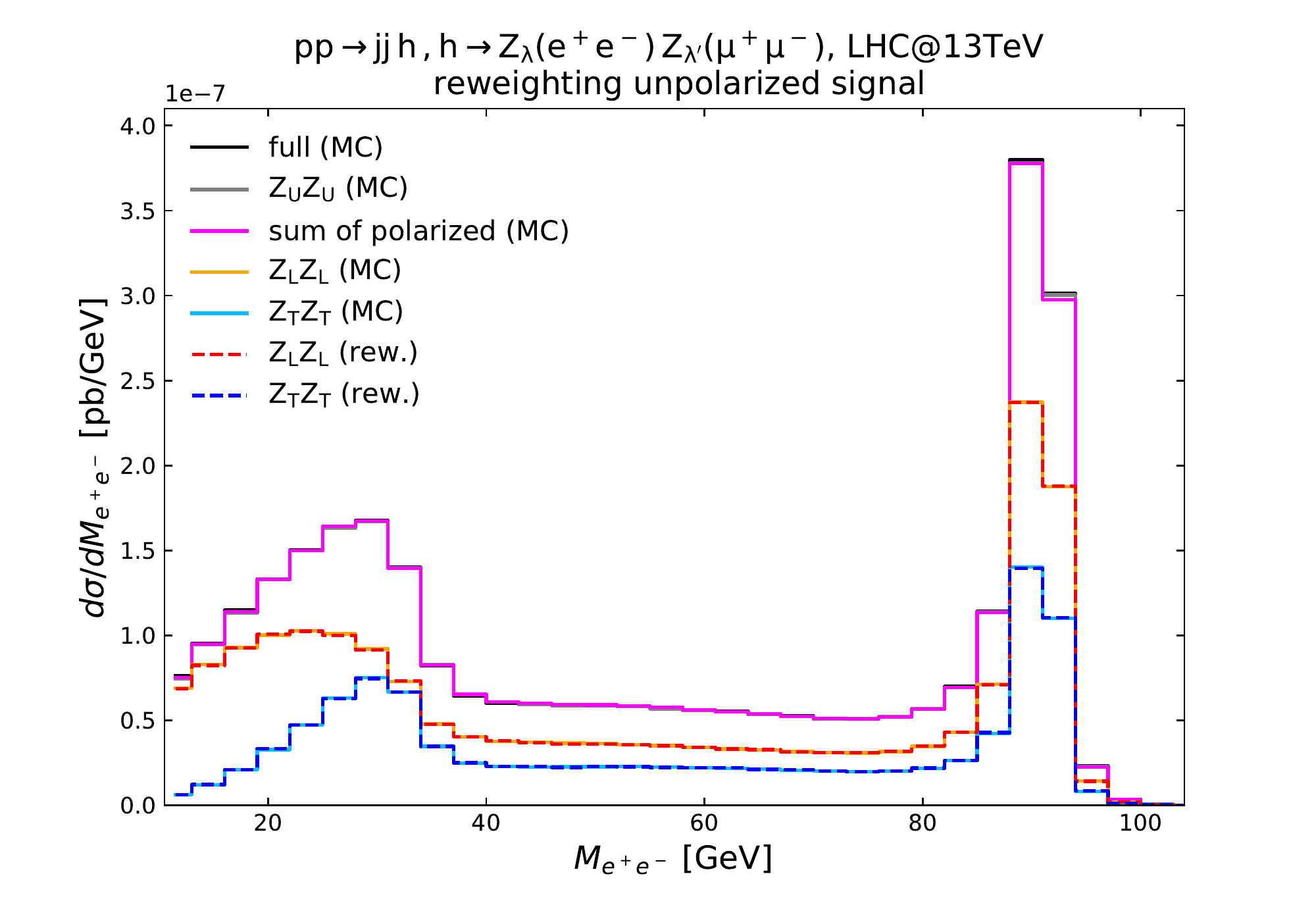}}
    \subfigure[$M_{\Pe^+\mu^+}$\label{meu_rew}]{\includegraphics[scale=0.44]{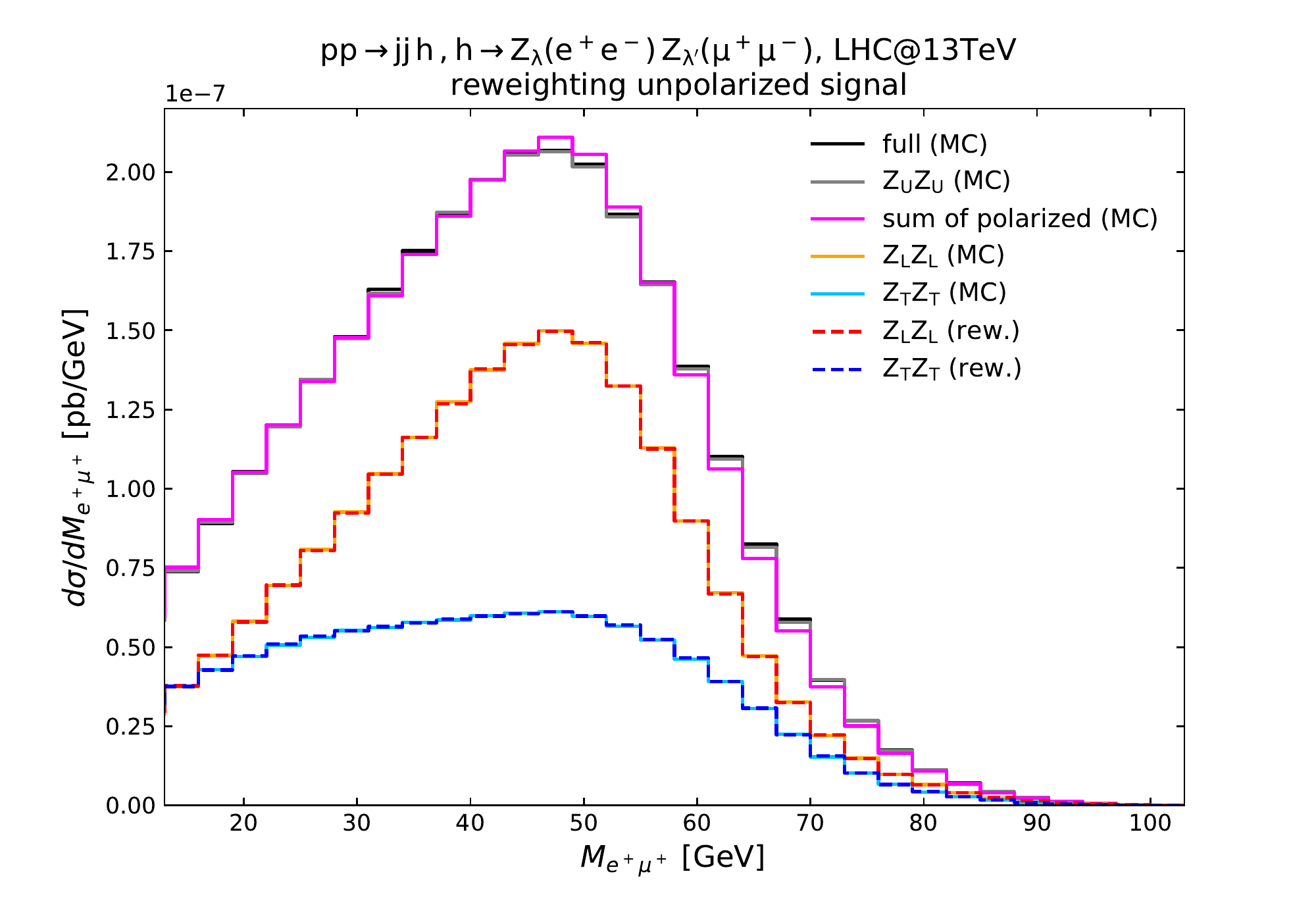}}   
    \caption{Differential cross-sections for VBF Higgs production at the LHC@13TeV:
      cosine of the positron decay angle in the corresponding $\PZ$ boson rest frame (a),
      azimuthal difference (in the Higgs rest frame) between the positron and the antimuon (b),
      invariant mass of the positron-electron pair (c)
      and of the positron-antimuon pair (d). Comparison between
      the Monte Carlo polarized distributions (solid curves, label: MC)
      and the results of the matrix-element reweighting of the
      unpolarized-signal distribution (dashed curves, label: rew.).
      The LEP setup described in Sect.~\ref{sec:setup} is understood.
    }\label{fig1rew}
  \end{figure*}

\section{Validation of the method}\label{sec:valid}
In order to validate the matrix-element-reweighting method (MERM), we first compare
the polarized cross sections computed with the MERM with those extracted from \phantommc in VBF.
In the first case, we reweight each unpolarized event as described in Sect.~\ref{reweighting}.
In the second case, we use the approach
of polarized amplitudes that has been already applied to VBS
and di-boson production \cite{Ballestrero:2017bxn,Ballestrero:2019qoy,Denner:2020bcz,Ballestrero:2020qgv,Denner:2020eck}.
Differently from previous studies in VBS
\cite{Ballestrero:2017bxn,Ballestrero:2019qoy,Ballestrero:2020qgv}, we do not apply any double-pole approximation (DPA),
since it is not possible to project both bosons on their mass shell. 
The comparison has been performed both in the INC setup and in the LEP one,
and the results are shown in Table~\ref{tab11}.

In the absence of lepton cuts, polarized cross sections can also be extracted from unpolarized
angular distributions of the charged leptons, via appropriate projections onto Legendre
polynomials \cite{Ballestrero:2017bxn}. The corresponding results are presented in the third
column of the INC section of  Table~\ref{tab11}.
They agree with those computed by other means at the sub-percent level. 
In the LEP setup the cross sections computed with the MERM and those obtained from
polarized amplitudes are in excellent agreement.
This check ensures that
separating polarizations of off-shell $\PZ$ bosons from Higgs decay is well defined.

The reweighting method works very well also at the differential level,
as can be appreciated in \reffi{fig1rew}, where we consider four kinematic
variables, in the LEP setup.
The polarized distributions are perfectly
reproduced by the reweighting both for the decay angle $\theta^*_{\Pe^+}$ [Fig.~\ref{thetastar_rew}]
that appears in Eqs.~\ref{ll}--\ref{pm}, and for the azimuthal difference $\Delta\phi_{\Pe^+\mu^+}$ computed
in the Higgs rest frame [Fig.~\ref{deltaphi_rew}] with the z-axis defined by the Higgs direction of flight
in the laboratory. Notice that this angular separation does not coincide with $\Delta\phi^*_{\Pe^+\mu^+}$
in Eqs.~\ref{ll}--\ref{pm}, where the z-axis is defined by the $\PZ$ boson direction of flight.
The invariant-mass variables considered in \reffis{mee_rew}--\ref{meu_rew}
are also perfectly described by the reweighting procedure.
We stress that the impressive agreement is motivated by the fact that the
reweighting factors of Eq.~\ref{eq:weight} are fully-differential. Therefore 
not only the variables appearing in the amplitude parametrization of Eq.~\ref{eq:master},
but any variable depending on the lepton momenta is well modelled, independently of
the selection cuts that are applied.

We have successfully validated the method also for other observables,
giving us confidence that the matrix-element reweighting procedure
furnishes a useful tool to generate polarized observables from unpolarized ones.

\section{Results}\label{sec:results}
In this section we only consider the LEP setup defined in Sect.~\ref{sec:setup}.

\begin{table}[tb]
\begin{center}
\renewcommand{\arraystretch}{1.3}
\begin{tabular}{c|cc}%
mode  & GGF &  VBF  \\[0.5ex] 
\hline
unp. & 6.988(5)$\cdot 10^{-6}\,$pb & 8.566(5)$\cdot 10^{-6}\,$pb\\[0.5ex]
\hline
LL  & 63.60\% & 64.26\% \\[0.5ex]
TT  & 36.35\% & 35.67\%\\[0.5ex]
interf.  & 0.05\% & 0.07\% 
\end{tabular}
\end{center}
\caption{Total cross-sections and polarization fractions in the LEP setup.
  Numerical errors for the unpolarized signals are shown in parentheses.
}\label{table:sigma}
\end{table}

\begin{figure*}[htb]
     \centering
     \subfigure[$\cos\theta^*_{\Pe^+}$\label{thetastar_rew_GF}]{\includegraphics[scale=0.4]{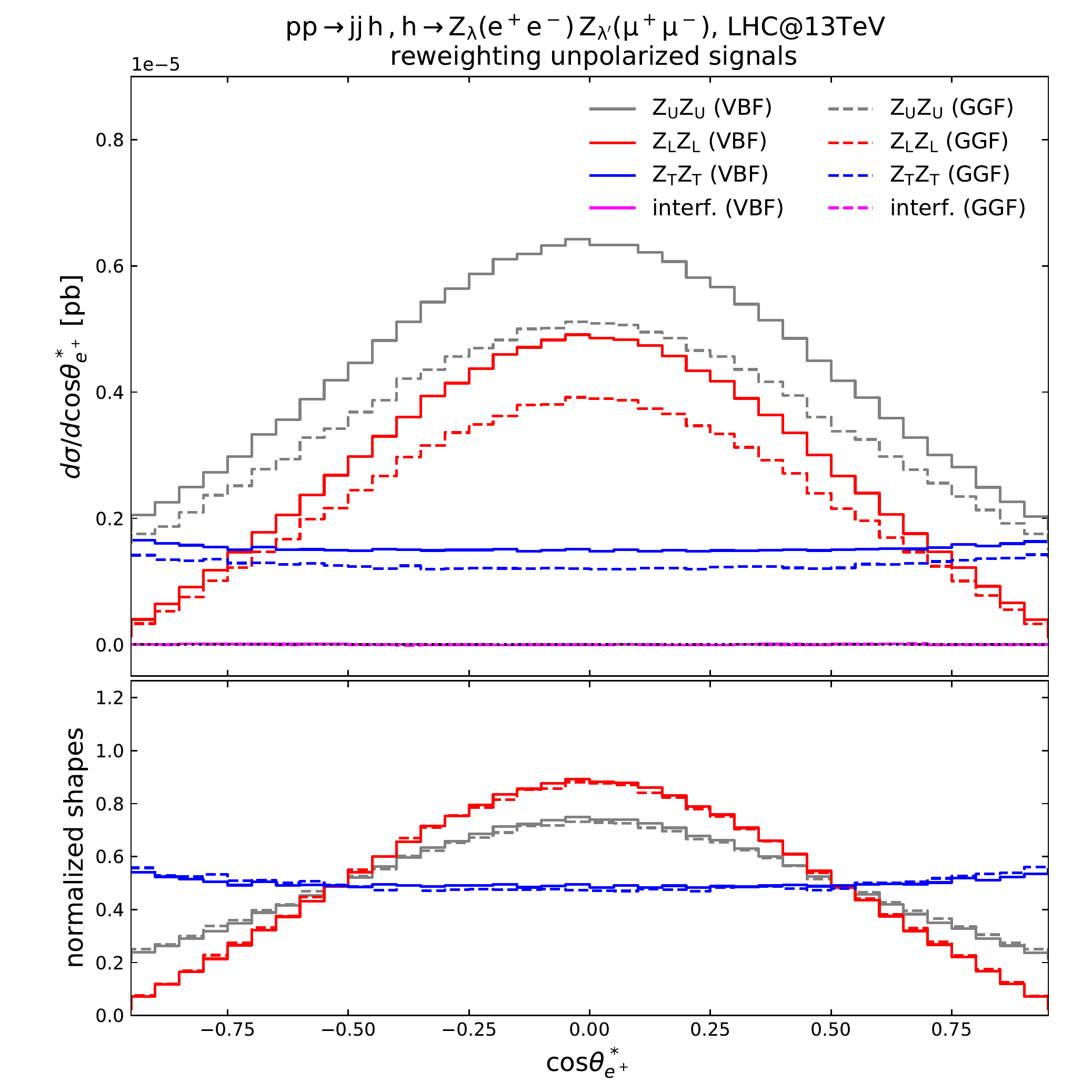}}
     \subfigure[$\Delta\phi_{\rm jj}$\label{dphijj_rew_GF}]{\includegraphics[scale=0.4]{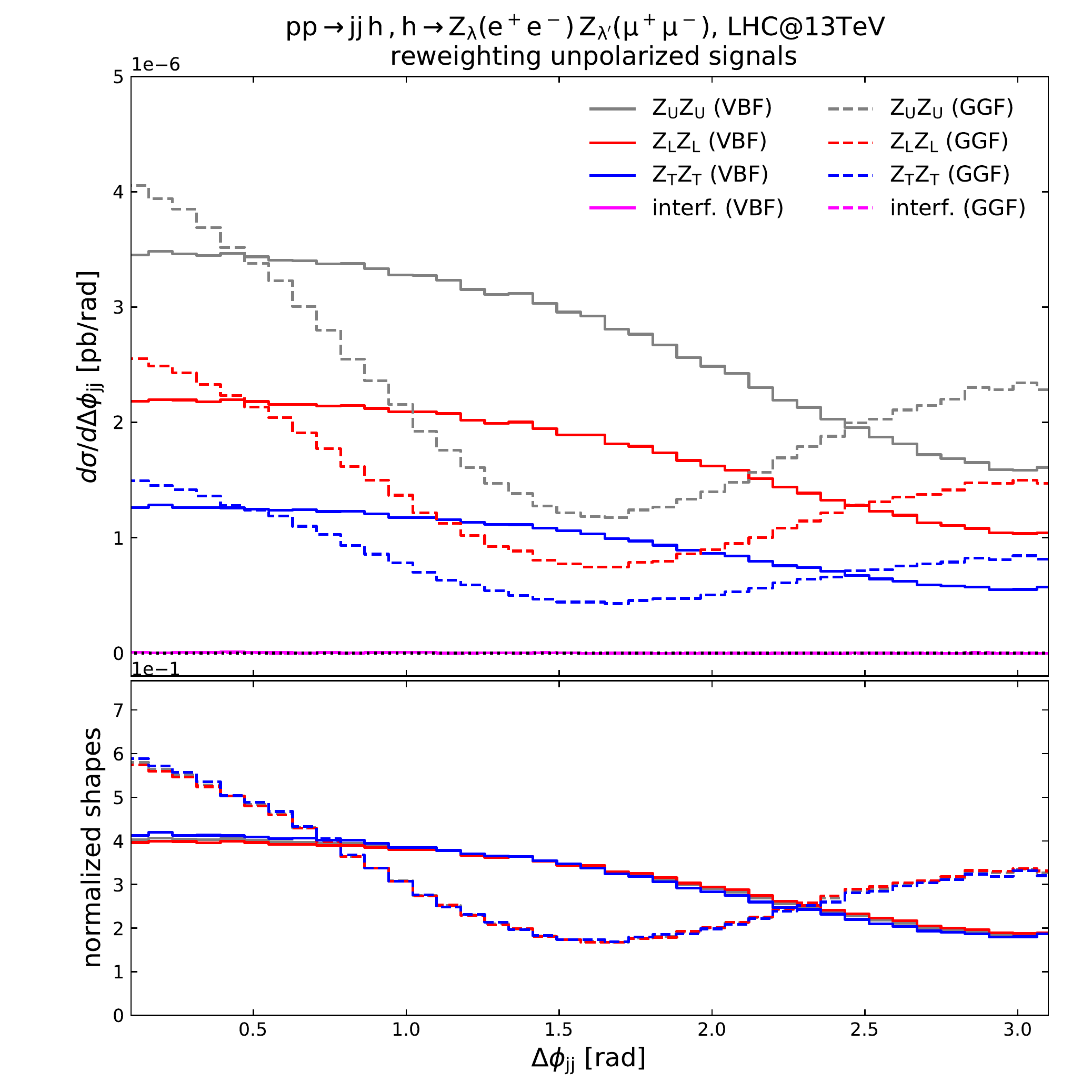}}
     \caption{Distributions for Higgs production in GGF (dashed curves) and VBF (solid curves) at the LHC@13TeV:
       cosine of the positron decay angle in the corresponding $\PZ$ boson rest frame (a) and azimuthal difference between the two jets (b).
       Polarized VBF (GGF) curves are obtained via the matrix-element reweighting of the unpolarized VBF (GGF) distribution.
       Top panel: differential cross-sections. Bottom panel: normalized distributions (unit integral). The LEP setup described
       in Sect.~\ref{sec:setup} is understood.
     }\label{fig1GF}
\end{figure*}

Before starting the discussion on the VBF and GGF signals, it is worth
commenting on the impact of some irreducible EW and QCD backgrounds.
The evaluation of such backgrounds to the Higgs signal is in fact
an important step in any polarization study.
We have computed with \phantommc the full process
\beq
\Pp\, \Pp \rightarrow \Pj\, \Pj \,\Pe^+\Pe^-\mu^+\mu^-\,,
\eeq
which, at LO order $[\mc O (\alpha^6)]$, receives contributions from several diagram topologies in which
no $s$-channel Higgs propagator is present.
{\color{black}However, the tight but realistic cut on the four-lepton invariant mass ($|M_{4\ell}-M_{\rm h}|<5\GeV$),
{\color{black}introduced in  sec\ref{sec:setup}}, suppresses these non-signal contributions.}
The contribution from Higgs-strahlung is suppressed by the large jet-pair invariant-mass cut.
The impact of all non-signal contributions on the total cross-section is below 0.5\%.

The fiducial cross-section of the QCD background, computed at LO $[\mc O (\as^2\alpha^4)]$ is 2.6\% of the VBF signal.
The signal cross-sections in GGF and VBF are shown in Table~\ref{table:sigma}.
{
The two LO cross-sections are of the same size.
Given the very low statistics of the process it is essential to sum over them,
provided that they show the same polarization structure.
The sizable scale uncertainties that characterize the GGF channel [$6.988(5)^{+37.7\%}_{-25.4\%}\cdot 10^{-6}$ pb]
make it important, in the lights of a realistic analysis, to include higher-order QCD corrections.
The LO scale uncertainties in VBF production are much smaller [$8.566(5)^{+3.9\%}_{-3.8\%}\cdot 10^{-6}$ pb] and the
NLO QCD corrections are expected to reduce them further.
As already pointed out, the MERM applies, in the
case of a Higgs-boson decay to four leptons, in exactly the same way 
whether radiative QCD corrections are included or not.
}

  Table~\ref{table:sigma} shows that the polarization fractions (polarized
  cross-sections over the unpolarized one)
  amount to 65\% for the longitudinal mode, 35\% for the transverse one.
  These fractions are
  almost the same for the two signals, showing that in the SM the polarization
  content only depends on the decay of the Higgs, and not on its production
  mechanism.
  The small differences in the fractions can be traced back to the slightly
    different kinematics for the two signals, which implies different effects
    of the selection cuts.
  The interferences are very small, in spite of quite tight lepton cuts.

  The results of Table~\ref{table:sigma} are not directly related to the SM Higgs coupling
  to vector bosons ($g_{\rm \stiny{\rm hVV}}$), which is independent of the polarization mode
  of the weak bosons. Therefore, a modification of $g_{\rm \stiny{\rm hVV}}$ due to new-physics
  effects would result in a enhanced (or diminished) cross-section at the unpolarized level,
  but one would not expect a strong modification of the relative weight of the longitudinal
  and the transverse mode.  
  However, as shown in \rf{Brehmer:2014pka}, if a new-physics model allows for
  different Higgs couplings to longitudinal and transverse vector bosons, the effect
  on the polarization fractions would be relevant. In addition, since the GGF cross-section
  depends on $g_{\rm \stiny{\rm hZZ}}^2$, while the VBF one depends either on $g_{\rm \stiny{\rm hZZ}}^4$
  or on $g_{\rm \stiny{\rm hZZ}}^2\,g_{\rm \stiny{\rm hWW}}^2$, polarization-dependent Higgs couplings
  to weak bosons could imply sizeable differences in the polarization structure
  of the GGF and the VBF signals.
  Therefore, a good understanding of both signals within the SM is the first step
  towards the search of beyond-SM effects via the polarization of weak-bosons.

  \begin{figure*}[htb]
     \centering
     \subfigure[$\Delta\phi^*_{\Pe^+\mu^+}$\label{dpeumstar_GF}]{\includegraphics[scale=0.4]{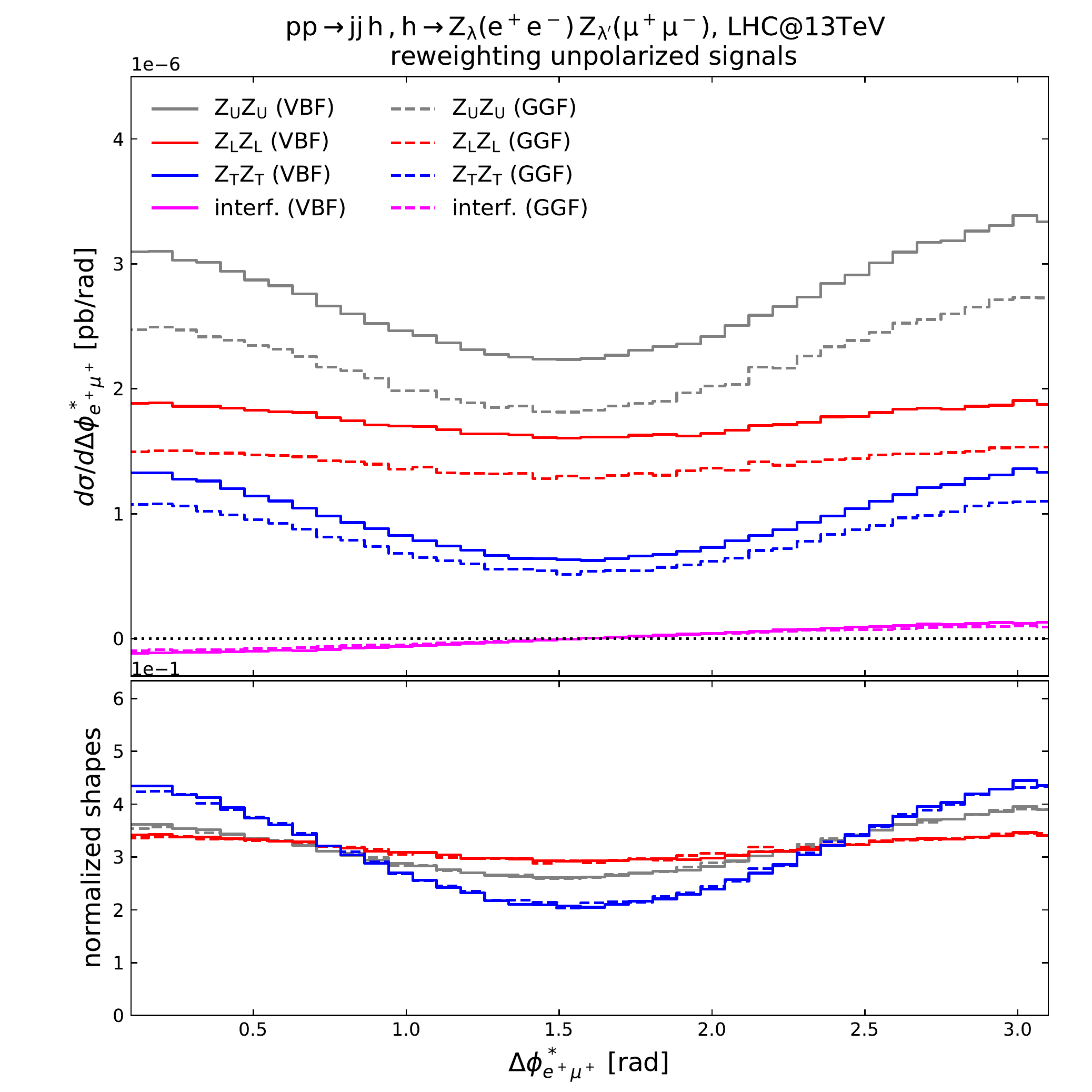}}
     \subfigure[$\Delta\phi_{\Pe^+\mu^+}$ \label{dpeup_GF}]{\includegraphics[scale=0.4]{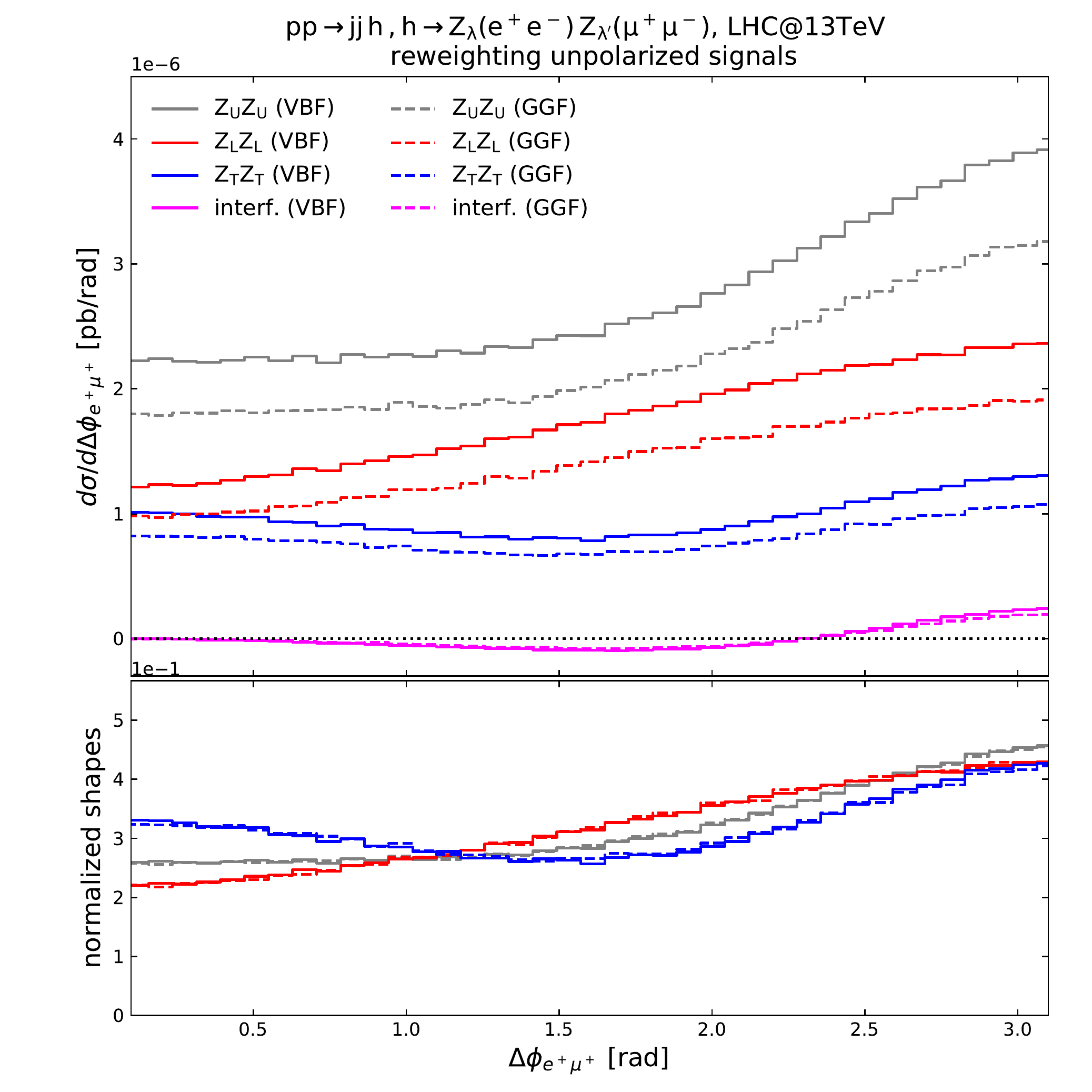}}
     \caption{Distributions for Higgs production in GGF (dashed curves) and VBF (solid curves) at the LHC@13TeV:
       difference between the azimuthal decay angles of the positron and of the antimuon, computed in the
       corresponding $\PZ$-boson rest frame, starting from the {Higgs-boson} rest frame (a),       
       azimuthal difference (in the Higgs rest frame) between the positron and the antimuon (b). Same structure as \reffi{fig1GF}.
     }\label{fig3GF}
   \end{figure*}

  The integrated results are not enough to fully characterize the
  polarization structure in the {Higgs-boson} decay.
  The study of differential distributions is essential to identify LHC observables
  that are capable of discriminating among polarization states.
  Furthermore, the interference terms could give noticeable shape distortions,
  in spite of an integrated contribution which is close to zero.

\begin{figure*}[htb]
     \centering
     \subfigure[$M_{\Pe^+\Pe^-}$\label{mee_GF}]{\includegraphics[scale=0.4]{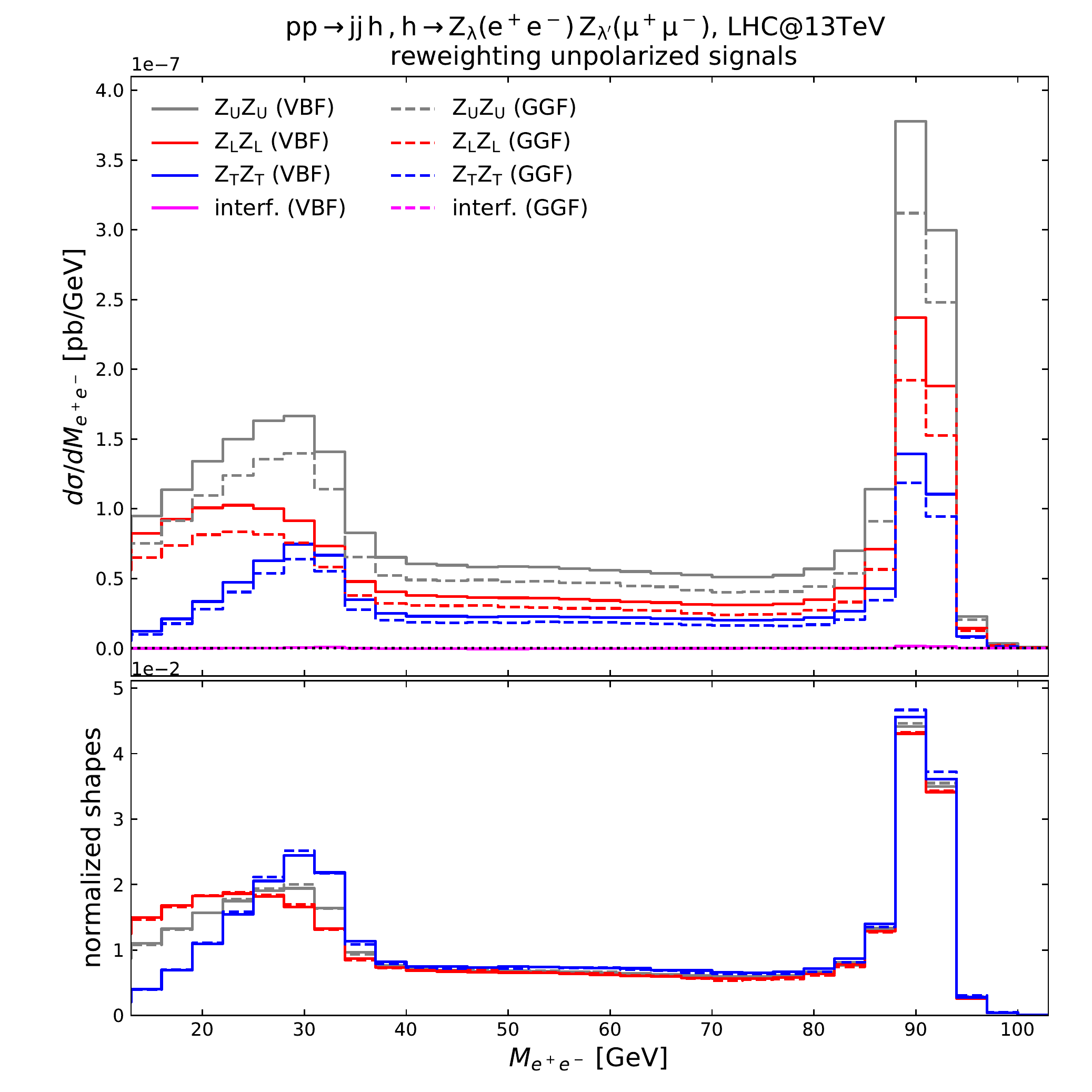}}
     \subfigure[$M_{\Pe^+\mu^+}$\label{meu_GF}]{\includegraphics[scale=0.4]{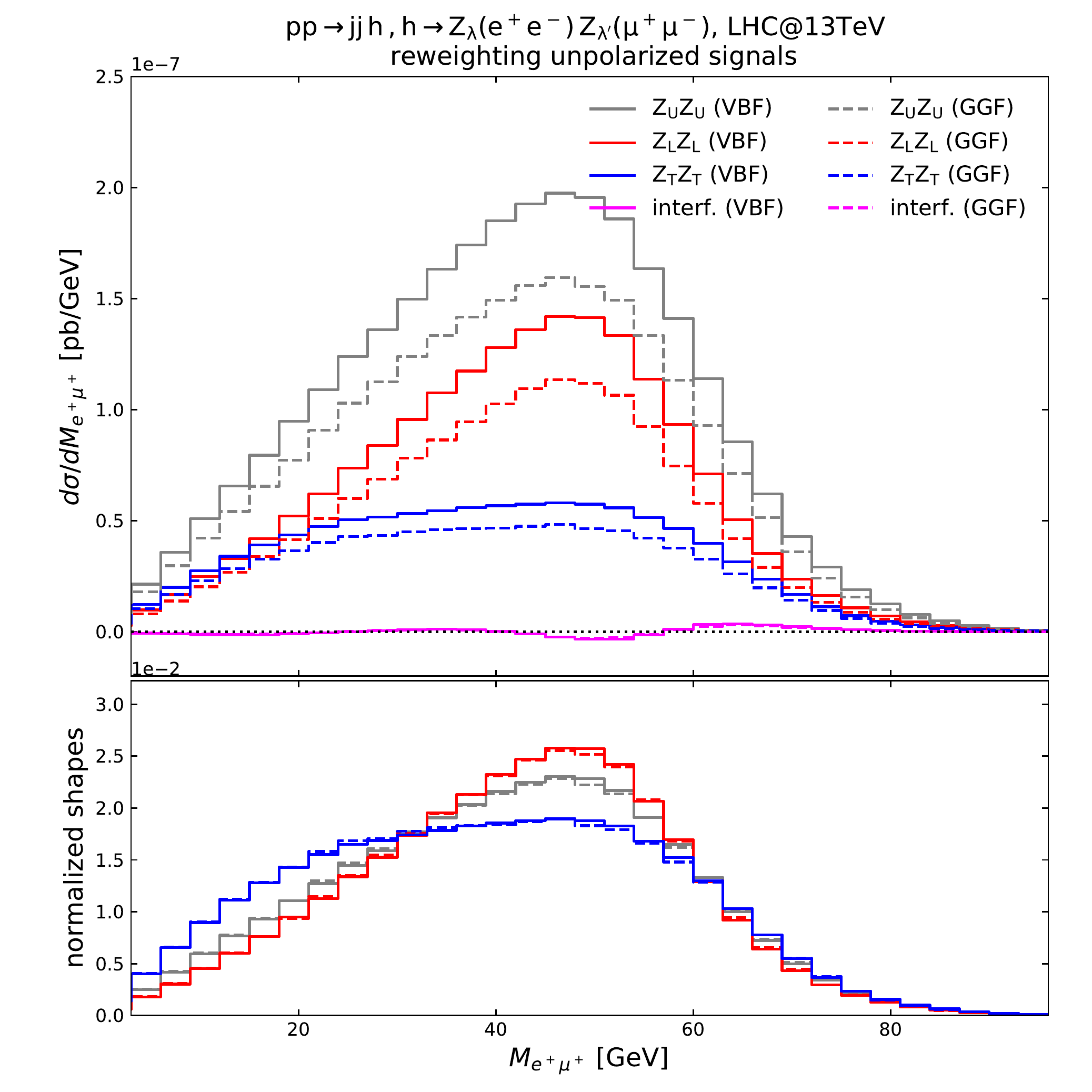}}
     \caption{Distributions for Higgs production in GGF (dashed curves) and VBF (solid curves) at the LHC@13TeV:
       invariant mass of the positron-electron pair (a) and of the positron-antimuon pair (b). Same structure as \reffi{fig1GF}.
     }\label{fig2GF}
   \end{figure*}

In \reffis{fig1GF}--\ref{fig2GF} we compare the distributions for different
polarization states, separating the GGF and VBF production mechanisms. We
present the results in terms of differential cross-sections (upper panels)
and normalized shapes (lower panels). 
  
  In \reffi{thetastar_rew_GF} we consider the cosine of the positron decay angle $\theta^*_{\Pe^+}$
  in the corresponding $\PZ$-boson rest frame, that is directly
  related to the polarization mode of the weak boson (see Eq.~\ref{eq:master}).
  This angular variable can be directly reconstructed at the LHC,
  thanks to the final state with four charged leptons.
  The VBF and GGF distributions have the same shape, both in the LL
  and in the TT component. The interferences play a negligible role for this
  distribution, as at the integrated level.  
  The symmetric LL shape features a maximum at $\cos\theta^*_{\Pe^+}=0$ and a
  minimum in the (anti)collinear regime, similar to the corresponding
  distribution in the INC setup ($d\sigma/d\cos\theta^*_{\Pe^+} \propto f_{\rL\rL}\,(1-\cos^2\theta^*_{\Pe^+})$).
  The TT distributions has constant convexity which is very close to zero,
  but with opposite sign w.r.t. the LL one. 
  The noticeable difference in shape between the LL and the TT
  distributions makes this variable well suited for polarization discrimination. 

  In \reffi{dphijj_rew_GF} we consider the azimuthal difference between the two
  tagging jets, computed in the laboratory frame.
  The GGF distribution shape is sizeably different from the VBF one, as expected
  from the different kinematics of the forward-backward jets in the two signals.
  In fact, this variable has been used to discriminate the GGF signal from its largest
  EW and QCD backgrounds in the $\rm h \rightarrow \PW\PW^*$ decay channel
  \cite{Klamke:2007cu}.
  The LL and TT distributions in a given signal do not show relevant differences,
  as the kinematics of the decay leptons is just mildly affected by the different
  kinematics of the production part of the amplitudes.
  This azimuthal difference can be useful for polarization measurements only
  in combination with other variables that discriminate among polarization modes,
  and if it is needed to separate different signal processes.

  In \reffi{fig3GF} we consider two azimuthal-difference variables that concern the
  charged-lepton kinematics.
  
  The difference between the two azimuthal decay angles of the positively-charged leptons
  $\Delta\phi^*_{\Pe^+\mu^+}$ is considered in \reffi{dpeumstar_GF}.
  This variable appears in the squared-amplitude parametrization of Eq.~\ref{eq:master},
  and gives a modulation to the interferences terms of Eqs.~\ref{lm}-\ref{pm}. 
  In the INC setup the LL distribution is flat, while the TT one it is
  dominated by a $\cos(2\Delta\phi^*_{\Pe^+\mu^+})$ functional dependence, since the interference
  between left and right modes is included in the cross-section \cite{Maina:2020rgd}. In the LEP setup, the
  behaviour of the TT distribution is quite similar to the one in the INC setup, while the
  LL shape is no more flat.
  The non-symmetric character of the unpolarized distributions is due to
  the longitudinal-transverse interferences that are negative for $\Delta\phi^*_{\Pe^+\mu^+}<\pi/2$,
  positive otherwise. This effect (at most of 2\%) reflects, even in the presence of lepton cuts, the
  $\cos(\Delta\phi^*_{\Pe^+\mu^+})$ functional dependence of the interference terms of Eqs.~\ref{lm}-\ref{lp}.
  The GGF and VBF signals show exactly the same behaviour in all polarized contributions and
  interference terms. The $\Delta\phi^*_{\Pe^+\mu^+}$ variable can be easily reconstructed
  at the LHC, with the considered final state, in the same fashion as $\cos\theta^*_{\Pe^+}$.

  In \reffi{dpeup_GF} we consider the azimuthal difference $\Delta\phi_{\Pe^+\mu^+}$
  between the positron and the antimuon computed in the {Higgs-boson} rest frame.
  This angular variable is related to $\Delta\phi^*_{\Pe^+\mu^+}$, with the
  difference that $\Delta\phi_{\Pe^+\mu^+}$ does not depend on angles computed in
  the $\PZ$-boson rest frame, but only on the kinematics of the four leptons in
  the {Higgs-boson} rest frame.
  The shapes of the polarized distributions is
  markedly different from  those for
    $\Delta\phi^*_{\Pe^+\mu^+}$.
  The interferences are very small and negative for $\Delta\phi_{\Pe^+\mu^+}<3\pi/4$,
  positive and slightly larger in size for $\Delta\phi_{\Pe^+\mu^+}>3\pi/4$.
  Also this variable enables a clear discrimination between the LL and TT modes.
  The LL shape is monotonically increasing from 0 to $\pi$, while the TT one
  has an absolute minimum at $\pi/2$, an absolute maximum at $\pi$ and another
  maximum at around 0.
  The impressive similarity of the SM polarized shapes for the GGF and VBF signals
  confirms that, if the focus is put on angular variables describing the {Higgs-boson}
  decay, it is safe to sum over production mechanisms.
  
  As pointed out in \rf{Maina:2020rgd}, not only angular variables but also
  invariant-mass observables are suitable for polarization discrimination.
  In \reffi{fig2GF} we consider the invariant mass of 
  the $\Pe^+\Pe^-$ and $\Pe^+\mu^+$ pairs.
  
  In the first case (\reffi{mee_GF}), the reconstructed $\PZ$-boson mass is peaked below $40\GeV$
  and at the pole mass, as at the Higgs resonance one $\PZ$ is typically on-shell while the other
  is far off-shell. Interferences are almost negligible for this observable.
  The difference between the LL and the TT normalized shapes concerns only the
  off-shell region below $40\GeV$: the longitudinal curve peaks between 20 and 25 GeV,
  while the transverse one has a narrower peak around 30 GeV.
  The GGF and VBF curves for a definite polarization state are almost identical,
  apart from mild differences in the TT curve around its two peaks.

  The invariant mass of the two positively-charged leptons is shown in \reffi{meu_GF}.
  The GGF and VBF signals behave in the same way, as for most of the other
  analyzed observables, giving distribution shapes that are almost independent
  of the production mode. The unpolarized and polarized distributions feature a
  maximum around $\Mz/2$, but the LL shape is narrower than the TT one,
  making this variable quite suitable for polarization discrimination.
  The polarization fractions (both for GGF and for VBF) show that the transverse
  mode gives a larger cross-sections than the longitudinal in the soft region
  ($M_{\Pe^+\mu^+}<20\GeV$), while the LL mode is larger in the
  rest of the spectrum.
  The interference pattern is slightly more evident than for $M_{\Pe^+\Pe^-}$. Its
  size is at most 2\% of the total at moderate masses ($M_{\Pe^+\mu^+}\approx 70\GeV$).
  Comparing \reffi{meu_GF} with Fig.~9 in \rf{Maina:2020rgd} one notices that
  different selection cuts can produce significant distortions in the observed distributions.
   
As a last comment, we note that in the SM not only the polarized shapes but also the
relative fraction of longitudinal and transverse modes are
almost independent of the production channel. This results in
unpolarized distributions shapes (gray curves in bottom panels of
\reffis{fig1GF}-\ref{fig2GF}) that are also independent of the production
channel.

\section{Conclusions}\label{sec:concl}
In this paper we have studied the polarization of $\PZ$ bosons decaying from
a Higgs boson produced in association with two jets, in a vector-boson-fusion kinematic regime.

We have proved that, thanks to the simple analytic structure of the Higgs-decay
Standard-Model amplitude, it is possible to avoid generating separately polarized event samples
by simply reweighting unpolarized events with fully-differential weights.

We have considered the two main channels that give contribution to Higgs+2j
production, namely gluon-gluon fusion and vector-boson fusion. The polarized
signals show the same behaviour in the two channels, both at the level of polarization
fractions, and at the level of the shape of polarized distributions. This allows to sum
over production channels. This holds for a SM measurement, and for modelling the background
for the search of beyond-the-Standard-Model effects.

The possibility of extending this work to higher-orders in perturbation theory
is also addressed. The usage of a single-pole approximation enables the
description of the Higgs decay to polarized bosons in the presence of EW radiative
corrections.

The results presented in this paper provide general techniques to study
polarizations of vector bosons from the decay of a scalar Higgs boson
produced in any channel at hadron colliders. The extension of such methods
to an effective-field-theory framework allows for a model-independent
assessment of the polarization structure at the Higgs resonance in the
presence of new-physics effects.

\section*{Acknowledgements}
We are grateful to Alessandro Ballestrero and Ansgar Denner
for useful and stimulating discussions.
EM is supported by the VBSCan COST Action CA16108 and by the SPIF 
(Precision Studies of Fundamental Interactions) INFN project.
GP is supported by the German Federal Ministry for
Education and Research (BMBF) under contract no.~05H18WWCA1.

\bibliographystyle{elsarticle-num}
\bibliography{higgs}

\end{document}